\documentclass{svjour3}

\smartqed

\usepackage{graphicx}
\usepackage{amsfonts}
\usepackage{mathrsfs}
\usepackage{amsmath}
\usepackage{epstopdf}

\begin{document}

\begin{center}
{\large {\bf {Modified Newtonian Gravity: Explaining observations of sub- and super-Chandrasekhar limiting mass white dwarfs}}}
\end{center}

Agrim Sharma\footnote{E-mail: agrimsharma@iisc.ac.in} and Banibrata Mukhopadhyay\footnote{E-mail: bm@iisc.ac.in} 

{\it Department of Physics, Indian Institute of Science, Bangalore 560012, India}

\vspace{1.0cm}

\noindent
{\it Abstract:} 
The idea of possible modification to gravity theory, whether it is in the 
Newtonian or general relativistic premises, is there for quite sometime.
Based on it, astrophysical and cosmological problems are targeted to
solve. But none of the Newtonian theories of modification has been performed
from the first principle. Here, we modify Poisson's equation and propose
two possible ways to modify the law gravitation which, however, reduces 
to Newton's law far away from the center of source. Based on these modified
Newton's laws, we attempt to solve problems lying with white dwarfs. There
are observational evidences for possible violation of the Chandrasekhar mass-limit 
significantly: it could be sub- as well as super-Chandrasekhar. We show that
modified Newton's law, either by modifying LHS or RHS of Poisson's equation,
can explain them.\\


\section*{1. Introduction}

Over the years, the researchers have explored the modifications to Einstein's
gravity in order to explain astrophysical and cosmological data. However, any
such modification proposed for the compact objects should be asymptotically
flat. It should follow the reduction from modified Einstein's to Einstein's
gravities and then to Newtonian gravity with distance from the source. 
Therefore, a modified Einstein's gravity may reduce to modified Newtonian gravity
at some length scale. 

In the present paper, we explore the possible modification to Poisson's equation
to understand the possible modification to Newtonian gravity. Based on that,
we target to resolve an astrophysical problem.

A white dwarf is a stellar core remnant composed mostly of electron-degenerate matter. The Chandrasekhar-limit is a theoretical limit for the maximum mass of a stable nonrotating and nonmagnetized white dwarf. If the mass of a white dwarf exceeds this limit, the force due to gravity becomes greater than that due 
to the electron degeneracy pressure. This leads to the star collapsing under its own gravity, leading to heating up of the plasma, which can result in a supernova. This is what happens in a type Ia supernova (SNIa).

A typical, slowly rotating, carbon-oxygen white dwarf accreting mass from a companion star explodes at a critical mass $\sim1.4M_{\odot}$. Due to this, 
all type Ia supernovae (SNeIa) have a characteristic light curve, that is luminosity as a function of time. This makes SNeIa a ``standard candle", which can be used to study the universe in various ways. Notably, observations of SNeIa led to the conclusion that the universe is undergoing an accelerated expansion.

However, there are observations of highly over-luminous SNeIa, e.g., SN~2003fg, SN~2006gz, SN~2007if, SN~2009dc~\cite{howell,scalzo} with progenitor masses believed to be as high as 2.8$M_{\odot}$ and highly under-luminous SNeIa, e.g., SN~1991bg, SN~1997cn, SN~1998de, SN~1999by, SN~2005bl with inferred progenitor masses being as low as 0.5 $M_{\odot}$~\cite{fillip,mazz,gon}.

As a possible explanation of these anomalous SNeIa, Mukhopadhyay and his collaborators had earlier explored modifications to general relativistic gravity models and the resulting change in the Chandrasekhar mass-limit~\cite{dm,km}.

This work explores similar modifications to Newton's model of gravity to see if modifications to corresponding classical quantities also produce similar results.

\section*{2. Modification to Newtonian gravity}
The aim here is to have a formula that is able to reproduce Newtonian results in the weak field limit and mimic general relativistic results in the strong field limit. The original Poisson's equation for Newtonian gravitational potential $\phi(\vec{r})$ for a density distribution $\rho(\vec{r})$ is
\begin{equation}
\boxed{\nabla^2\phi=4\pi\rho G}.
\end{equation}

The following are two general types of modifications that are considered in this work:
\begin{equation}
\boxed{\nabla^2\phi+A\nabla^4\phi=4\pi\rho G}
	\label{poileft}
\end{equation}
\begin{center}
and
\end{center}
\begin{equation}
\boxed{\nabla^2\phi=4\pi G(\rho+B\rho^2+...)}.
\end{equation}

\section*{3. LHS modification: general solution to the equation}
 The modified formula can be provided as 
\begin{equation}
\nabla^2\phi+A\nabla^4\phi=4\pi\rho G.
	\label{poileft2}
\end{equation}

Using Green's method, with the constraint that $\phi \in \mathbb{R}$
\begin{equation}
\text{for $A>$0: }\boxed{\phi(\vec{r}) = \int \frac{d^3\vec{r'}}{4\pi |\vec{r}-\vec{r'}|} \left(\int d^3 \vec{r''} \frac{\cos \left({\frac{|\vec{r'}-\vec{r''}|}{A}}\right)}{4\pi |\vec{r'}-\vec{r''}|} \frac{4\pi G \rho(\vec{r''})}{A}\right)},
\end{equation}

\begin{equation}
\text{for $A<$0:} \boxed{\phi(\vec{r}) = \int \frac{d^3\vec{r'}}{4\pi |\vec{r}-\vec{r'}|} \left(\int d^3 \vec{r''} \frac{\exp \left({-\frac{|\vec{r'}-\vec{r''}|}{\sqrt{|A|}}}\right)}{4\pi |\vec{r'}-\vec{r''}|} \frac{4\pi G \rho(\vec{r''})}{|A|}\right)}.
\end{equation}

We have provided detailed explanation in Appendix A.

\subsection*{3.1 Calculation of the mass-limit}
Coming to the ideal white dwarf model which satisfies the equations:
\begin{equation}\label{Pvsrho}
P=K\rho^{1+\frac{1}{n}}, 
\end{equation}

\begin{equation}\label{Mvsrho}
	\frac{dM_r}{dr}=4\pi r^2\rho,
\end{equation}

\begin{equation}\label{Pvsphi}
	\frac{\nabla P}{\rho}=-\nabla \phi,
\end{equation}

\begin{equation}\label{P'vsphi}
\implies\frac{1}{r^2}\frac{d}{dr}(\frac{r^2}{\rho} \frac{dP}{dr})=-\nabla^2 \phi =-f(r),
\end{equation}
where $P$ is the pressure of white dwarf matter, $\rho$ the density, $M_r$ the
mass enclosed in the radius $r$, $n$ the polytropic index and $K$ the 
polytropic constant.

Let $\theta$ be a dimensionless function of $r$ so that
\begin{equation}
\rho(r)\equiv\rho=\rho_c\theta^n,
\end{equation}
where $\rho_c$ being the density at the centre of the white dwarf.

Similarly, considering dimensionless variable $\xi$ such that
\begin{equation}
r=a\xi,
\end{equation}
we obtain
\begin{equation}
-f(r)
=\frac{1}{r^2}\frac{d}{dr}\left (\frac{r^2}{\rho} \frac{dP}{dr}\right )
=\frac{(1+n)K\rho_c^{1/n}}{a^2}\frac{1}{\xi^2}\frac{d}{d\xi}\left (\xi^2 \frac{d\theta}{d\xi}\right ).
\end{equation}

From the modified equation for gravity given by Eq. (\ref{poileft2}):
\begin{equation}
f(r)+A \frac{1}{r^2}\frac{d}{dr}\left (r^2 \frac{df}{dr}\right ) = g(r)=4\pi\rho_c\theta^nG,
\end{equation}

\begin{equation}
\left (\frac{2}{\xi} \frac{d\theta}{d\xi} + \frac{d^2\theta}{d\xi^2}\right )
+\frac{A}{a^2} \left (\frac{4}{\xi} \frac{d^3\theta}{d\xi^3} + \frac{d^4\theta}{d\xi^4}\right ) =-\frac{a^2 4\pi G}{(1+n)K\rho_c^{\frac{1-n}{n}}}\theta^n.
\end{equation}

If we let $a^2=\frac{(1+n)K\rho_c^{\frac{1-n}{n}}}{4\pi G}$, then we obtain
\begin{equation} \label{laneEmden}
\left (\frac{2}{\xi} \frac{d\theta}{d\xi} + \frac{d^2\theta}{d\xi^2}\right )
+\frac{A}{a^2} \left (\frac{4}{\xi} \frac{d^3\theta}{d\xi^3} + \frac{d^4\theta}{d\xi^4}\right ) =-\theta^n.
\end{equation}

Notice that when $A=0$, the above equation reduces to the Lane-Emden equation as expected.

For $n=3$ (i.e. the relativistic case), we will numerically find the solution of $\theta(\xi)$, which will give us the mass-limit of a white dwarf as a function of $A/a^2$ based on
\begin{equation}
M=\int_0^R 4\pi r^2\rho dr=4\pi a^3 \rho_c \int_0^{\xi_1} \xi^2 \theta^n d\xi.
\end{equation}

\subsection*{3.2 Numerical solution of $\rho(r)$ for white dwarfs}
Having added a new parameter to the Lane-Emden equations, we will need some extra information than what is usually needed to find numerical solutions in the
Newtonian gravity.
Fig. 1 shows the variation of $M_{lim}/M_{Ch}$, with $M_{lim}$ being 
new mass-limit and $M_{Ch}$ being original Chandrasekhar-limit, as a function of $a^2/A$ for different values of the extra unknown constraint $\theta''(0)$ shown as labels.

Since `$a$' depends on $\rho_c$ and $n$, we can obtain different values of Chandrasekhar-limit for different $\rho_c$ and even subtly different $n$, both of 
which are physical parameters. Hence given a fixed value of the modification parameter $A$ which we naturally expect to be a universal constant, there can be conditions where the Chandrasekhar mass-limit differs depending on conditions inside the white dwarf.

\begin{figure}[h]
	\makebox[\textwidth][c]
	{\includegraphics[width=1.2\textwidth]{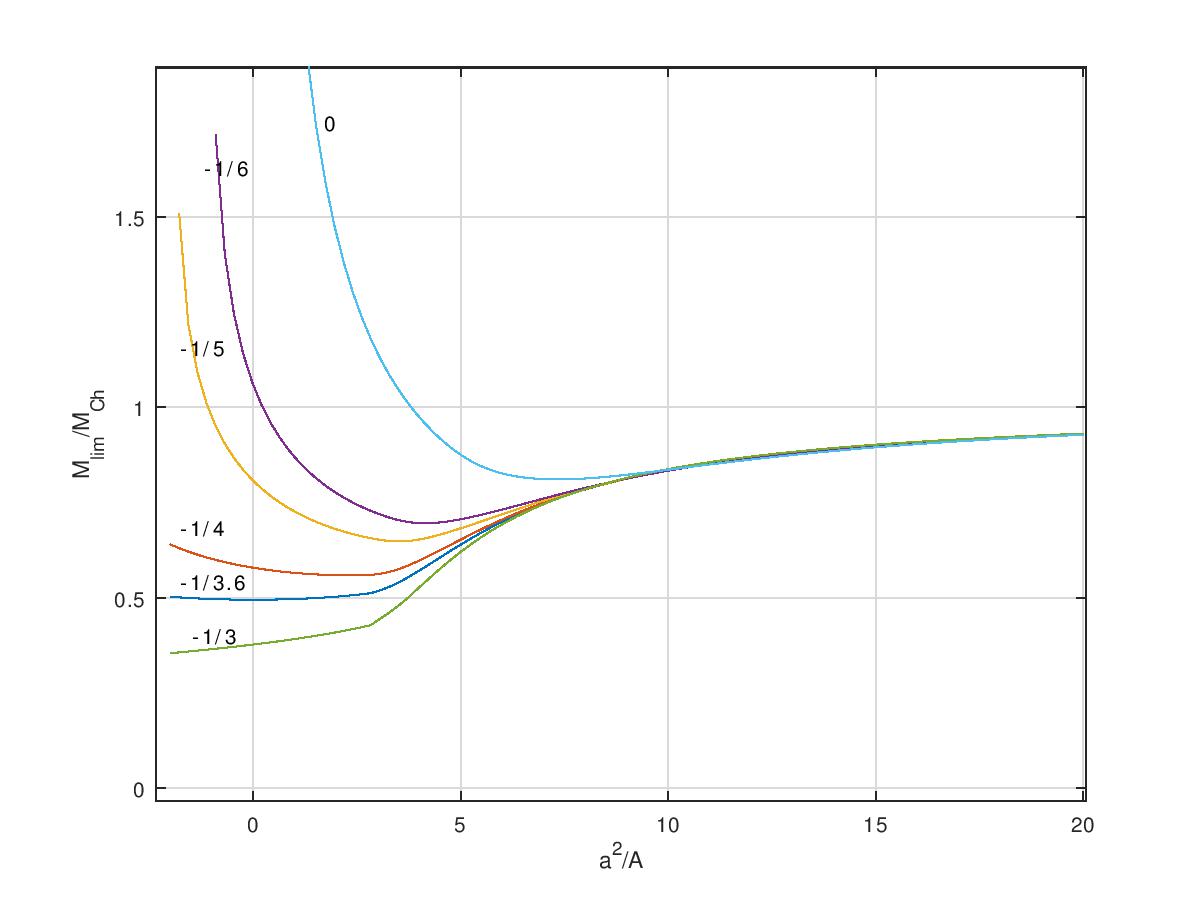}}	
	\caption{Variation of $M_{lim}/M_{Ch}$ with $a^2/A$ for different $\theta''(0)$ shown as labels.}
\end{figure}

\subsection*{3.3 Analytical limits of $M_{lim}/M_{Ch}$ as $A\to 0$ and $A\to\infty$}
The modified Lane-Emden equation is:
\begin{equation} 
\left (\frac{2}{\xi} \frac{d\theta}{d\xi} + \frac{d^2\theta}{d\xi^2}\right )
+\frac{A}{a^2} \left (\frac{4}{\xi} \frac{d^3\theta}{d\xi^3} + \frac{d^4\theta}{d\xi^4}\right ) =-\theta^n.
\end{equation}

Analytical limits can give us some verification of the correctness of numerical solutions. For example, the results in Fig. 1 appear to be correct atleast in the $a^2/A\geq0$ regime.

\subsubsection*{3.3.1 Limit $A\to 0$:}
For $A/a^2 << 1$, assuming slight perturbation to the original Lane-Emden equation, the solution gives us:
\begin{equation}
\frac{M_{lim}}{M_{Ch}}=\frac{\int_0^{\xi_1} \xi^2 \theta^3 d\xi}{2.01824}
= \Bigg(1 + \frac{A}{a^2}\frac{  2.1173}{2.01824}
\Bigg)
\Big(1-3\frac{A}{a^2}\Big).
\end{equation}
Fig. 2 confirms that the solution passes $A=0$ point smoothly with $M_{lim}/M_{Ch}=1$,
confirming the correctness of results around $A=0$.

\begin{figure}[h]
\centering
	\includegraphics[width=\textwidth]{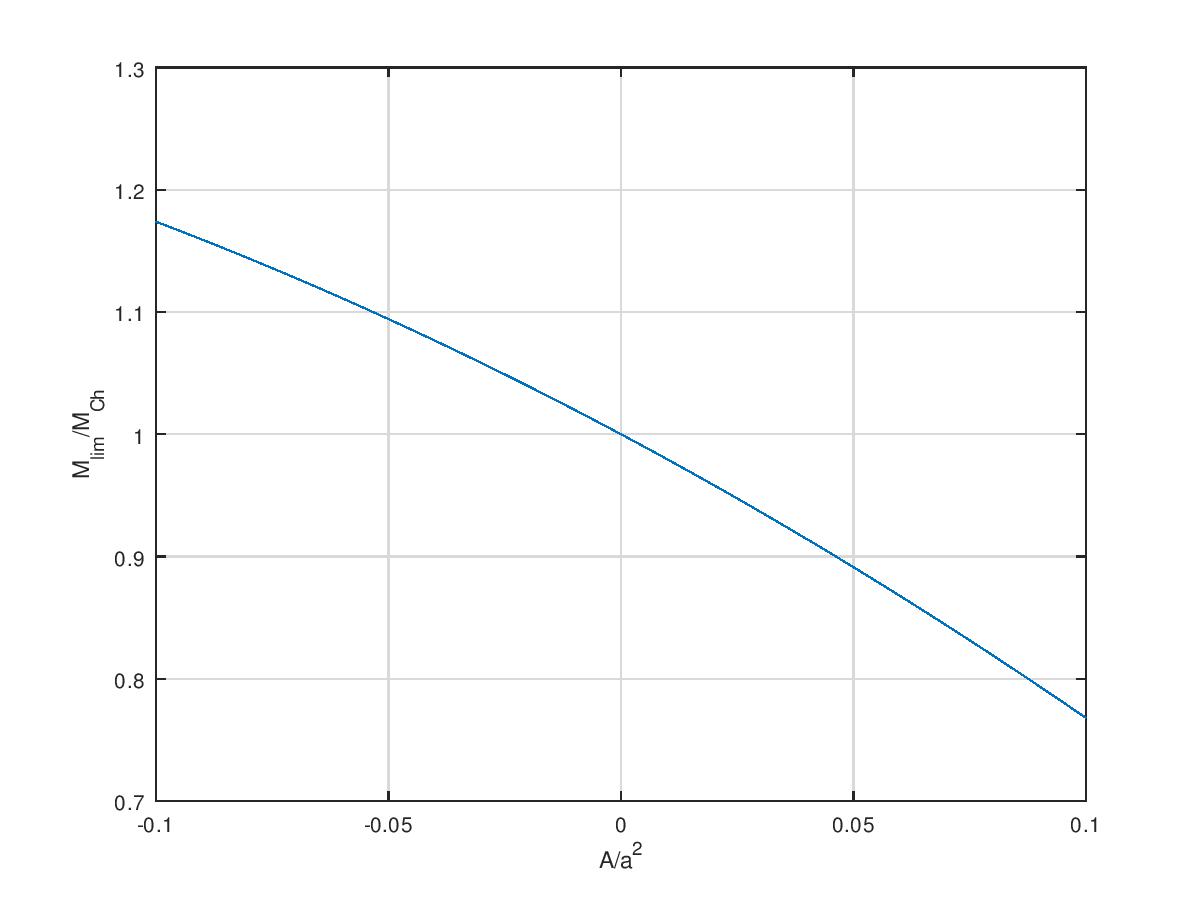} 
	\caption{Variation of $M_{lim}/M_{Ch}$ with $A/a^2$ around $A=0$.}
\end{figure}

\subsubsection*{3.3.2 Limit $A\to \infty$:}
In this case the mass-limit depends on second derivative of 
$\theta(r)=\left[\rho(r)/\rho(0)\right]^{1/n}$ (where $\rho(0)=\rho_c$) at the centre of the white dwarf, and hence we obtain
\begin{equation}
	\frac{M_{lim}}{M_{Ch}}=\frac{1}{15}\Bigg( \frac{-2}{\theta''(0)}\Bigg)^{3/2}.
\end{equation}

\section*{3.4 Mass-radius relation}\label{MR_LHS}
For the mass-radius relation, the density as a function of distance from center is numerically computed\footnote{Using GNU Octave's 'ode15s' function - a variable step, variable order method based on Backward Difference Formulas (BDF).} using the following equations:
\begin{equation}
	\frac{dM_r}{dr}=4\pi r^2\rho,
\end{equation}

\begin{equation}
	\frac{\nabla P}{\rho}=-\nabla \phi.
\end{equation}

Chandrasekhar's exact equation of state is given by
\begin{equation}
P = K_1 \left[ x(2x^2-3) \sqrt{x^2+1} + 3 \sinh^{-1} x \right],
\end{equation}

\begin{equation}
\rho = K_2 x^3,
\end{equation}
where $K_1=8\pi \mu_e m_H (m_e c)^3/3h^3$ and $K_2=\pi m_e^4 c^5/3h^3$.

The modified gravity equation is:
\begin{equation}
\nabla^2\phi+A\nabla^4\phi=4\pi\rho G.
\end{equation}

Fig. 3 shows the variation radius of white dwarfs as a function of their mass.

\begin{figure}[h]
\centering
	\includegraphics[width=0.8\textwidth]{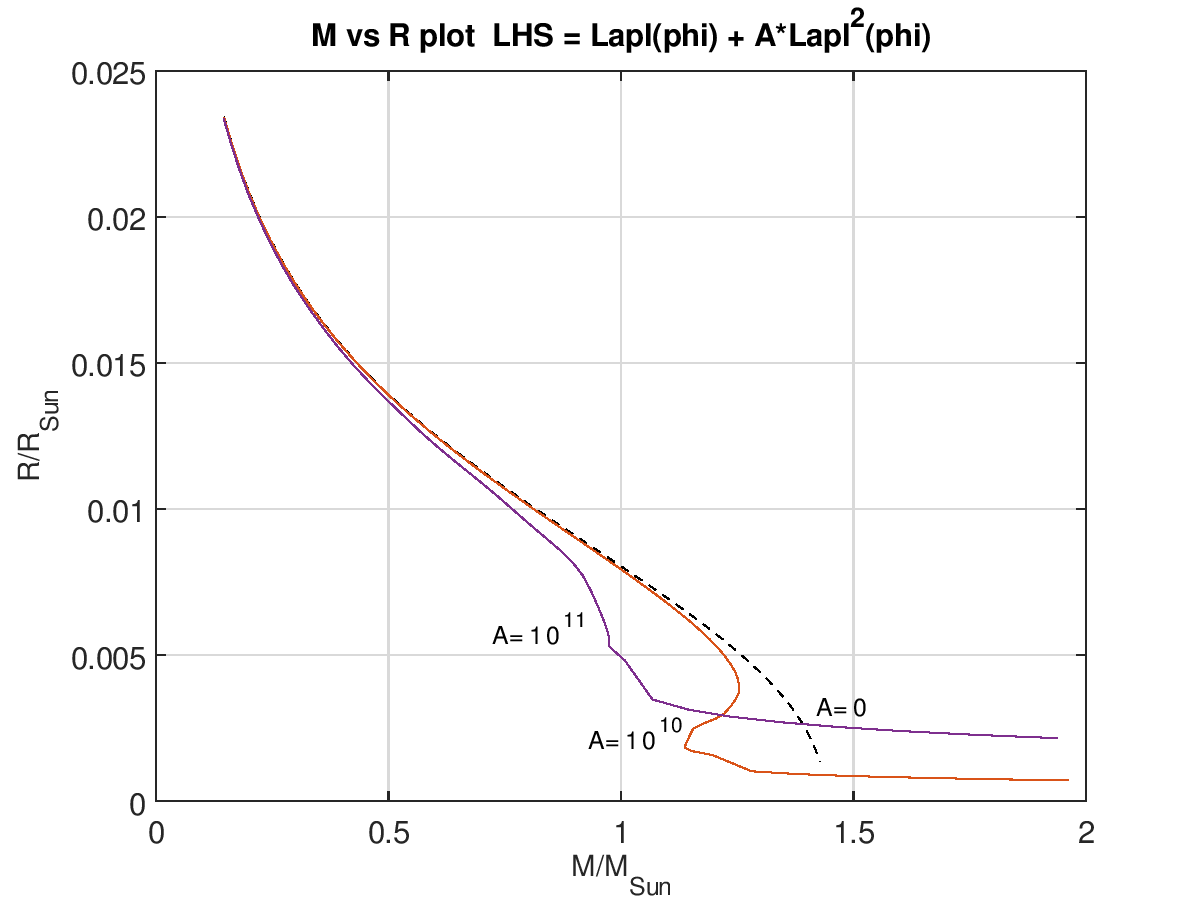} 
	\caption{Variation of radius with mass, where $A$ is in units of $m^2$. The central density varies from $\rho_c=10^{8}$ to $10^{13} kg/m^3$.}
\end{figure}

\section*{4. RHS modification: general solution to the equation}
The modified formula in this case is
\begin{equation}
\nabla^2\phi=4\pi G(\rho+B\rho^2+...).
\end{equation}

Let $\rho_{\rm eff}=\rho+B\rho^2+...$ 

We explicitly know the solution of the Poisson equation: $\nabla^2\phi=\rho_{\text{eff}}(\vec{r})$ is
\begin{equation}
\boxed{\phi(\vec{r})=- \int \frac{\rho_{\rm eff}(\vec{r'})}{4\pi|\vec{r}-\vec{r'}|} d^3\vec{r'}.}
\end{equation}
Therefore, the form of gravitational potential will be exactly same as the usual Newtonian gravitational potentials, but with $\rho_{\rm eff}$ instead of $\rho$.

\subsection*{4.1 Mass-radius relation}
Using the same method as in section 3, with the new modified gravity equation, the variation of radius for white dwarfs as a function of their mass 
is generated numerically\footnote{Using GNU Octave's 'ode15s' function - a variable step, variable order method based on Backward Difference Formulas (BDF).}.
The modified equation considered is:
\begin{equation}
\nabla^2\phi=4\pi G(\rho+A\rho^2)
\end{equation}

\begin{figure}[h]
\centering
	\includegraphics[width=11cm]{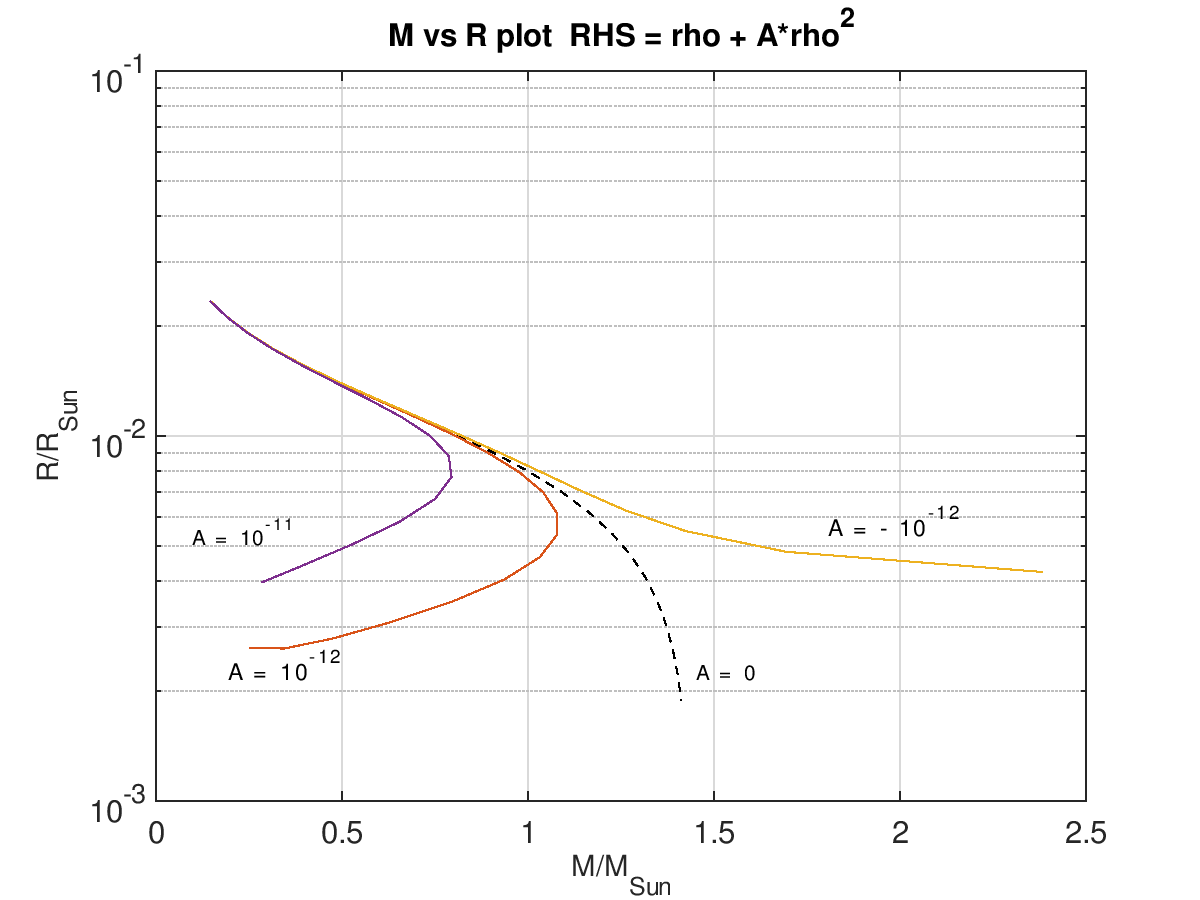} 
	\caption{Variation of radius with mass, where $A$ is in units of $m^3/kg$. The central density varies from $\rho_c=10^{8}$ to $10^{11} kg/m^3$.}
\end{figure}

One can observe in Fig. 4 that when $\rho_{\rm eff}>\rho$, the mass is lower 
for a given radius, as expected, and likewise for $\rho_{\rm eff}<\rho$, the mass is higher than usual white dwarf mass, obtained based on Newton's law, 
for the same radius. Therefore, using higher order polynomial terms, at different densities, the white dwarf can show arbitrarily small or large masses in the range of $\rho$ where the respective terms dominate.

\section*{5. Conclusion}
Newton's law is a remarkably successful physics, well tested in laboratory, 
also is remarkably successful in explaining low energy physics. Several
astrophysical features are also quite abide it. In this connection,
the Chandrasekhar-limit perhaps is one of the most celebrated astrophysical 
discoveries in the 20th century, whose physical insight can be well understood
in the Newtonian framework itself. However, observations of several
peculiar over- and under-luminous SNeIa for about last three decades argue for the significant
violation of the Chandrasekhar mass-limit. We have shown here that appropriate
modifications to Poisson's equation and, hence, Newton's law can explain the significant violation of the 
Chandrasekhar-limit, as inferred from observations. It argues that while
the existence of the Chandrasekhar-limit is sacrosanct, its value need not be.
We expect the proposed modifications to Poisson's equation and modified 
Newton's law, which reduces to Newton's law asymptotically, to have far reaching implications.

\section*{Appendix A: General solution of the equation with modified LHS}
The modified equation is given by
\begin{equation}
\nabla^2\phi+A\nabla^4\phi=4\pi\rho G.
\end{equation}
$$\text{Let  }u(\vec{r}) = \nabla^2\phi(\vec{r}),$$
\begin{equation}
\implies \left[A\nabla^2+1\right] u(\vec{r}) = 4\pi\rho G.
\end{equation}
From the solution of Screened Poisson Equation (derived using Green's functions)
\begin{equation}
\left[\nabla^2-\lambda^2\right] u(\vec{r}) = -f(\vec{r}),
\end{equation}
\begin{equation}
\implies u(\vec{r})=\int d^3 \vec{r'} \frac{e^{-\sqrt{\lambda^2} |\vec{r}-\vec{r'}|}}{4\pi |\vec{r}-\vec{r'}|} f(\vec{r'}).
\end{equation}

For $A<0$,\hspace{2.5cm} $\lambda^2 = \frac{1}{|A|}, f =  \frac{4\pi\rho G}{|A|}$,
$$u(r\to\infty)\to 0 \implies \lambda = \frac{1}{\sqrt{|A|}}.$$
Hence,
\begin{equation}
u(\vec{r}) = \nabla^2\phi(\vec{r})\implies \phi(\vec{r}) = \int \frac{d^3\vec{r'}}{4\pi |\vec{r}-\vec{r'}|}u(\vec{r'}), 
\end{equation}
\begin{equation}
\implies \boxed{\phi(\vec{r}) = \int \frac{d^3\vec{r'}}{4\pi |\vec{r}-\vec{r'}|} \left(\int d^3 \vec{r''} \frac{\exp \left({-\frac{|\vec{r'}-\vec{r''}|}{\sqrt{|A|}}}\right)}{4\pi |\vec{r'}-\vec{r''}|} \frac{4\pi G \rho(\vec{r''})}{|A|}\right).}
\end{equation}

Similarly, for $A>0$,\hspace{1cm} $\lambda^2 = -\frac{1}{A}, f =  -\frac{4\pi\rho G}{A}$,
$$u(r)\in \mathbb{R} \implies e^{-\lambda|\vec{r}-\vec{r'}|} \equiv \frac{1}{2}\left(e^{\frac{-\iota}{\sqrt{A}}|\vec{r}-\vec{r'}|}+ e^{\frac{\iota}{\sqrt{A}}|\vec{r}-\vec{r'}|}\right) \\ \equiv \cos\left(\frac{1}{\sqrt{A}}|\vec{r}-\vec{r'}|\right).$$
Therefore,
\begin{equation}
\implies \boxed{\phi(\vec{r}) = \int \frac{d^3\vec{r'}}{4\pi |\vec{r}-\vec{r'}|} \left(\int d^3 \vec{r''} \frac{\cos \left({\frac{|\vec{r'}-\vec{r''}|}{A}}\right)}{4\pi |\vec{r'}-\vec{r''}|} \frac{4\pi G \rho(\vec{r''})}{A}\right).}
\end{equation}

\bibliographystyle{unsrt}

\begin{thebibliography}{99}

\bibitem{howell} D. A. Howell et al., {\it Nature} \textbf{443} (2006) 308.
\bibitem{scalzo} R. A. Scalzo et al., {\it ApJ} \textbf{713} (2010) 1073.
\bibitem{fillip} A. V. Filippenko et al., {\it AJ} \textbf{104} (1992) 1543.
\bibitem{mazz} P. A. Mazzali et al., {\it Mon. Not. R. Astron. Soc.} \textbf{284} (1997) 151.
\bibitem{gon} S. Gonz\'alez-Gait\'an et al., {\it ApJ} \textbf{727} (2011) 107.
\bibitem{dm} U. Das and B. Mukhopadhyay, {\it JCAP} \textbf{5} (2015) 045.
\bibitem{km} S. Kalita and B. Mukhopadhyay, {\it JCAP} \textbf{9} (2018) 007.


\end{thebibliography}

\end{document}